\documentclass[conference]{IEEEtran}
\usepackage{graphicx}
\usepackage{cite}
\usepackage{url}

\begin{document}

\title{Control Systems: an Application to\\
a High Energy Physics Experiment (COMPASS)}

\author{
    \IEEEauthorblockN{P.\ Bordalo\IEEEauthorrefmark{2}\IEEEauthorrefmark{3}, A.S.\ Nunes\IEEEauthorrefmark{2}\IEEEauthorrefmark{1}, C.\ Pires\IEEEauthorrefmark{2}, C.\ Quintans\IEEEauthorrefmark{2} and S.\ Ramos\IEEEauthorrefmark{2}\IEEEauthorrefmark{3}}
    \IEEEauthorblockA{\IEEEauthorrefmark{2}LIP, 1000-149 Lisbon, Portugal}
    \IEEEauthorblockA{\IEEEauthorrefmark{3}IST, Universidade T\'ecnica de Lisboa, Lisbon, Portugal}
    \IEEEauthorblockA{\IEEEauthorrefmark{1}Corresponding author. {\it E-mail address}: Ana.Sofia.Nunes@cern.ch}
}

\maketitle

\begin{abstract}
The Detector Control System (DCS) of the COMPASS experiment at CERN is presented. The experiment has a high level of complexity and flexibility and a long time of operation, that constitute a challenge for its full monitorisation and control. A strategy to use a limited number of standardised, cost-effective, industrial solutions of hardware and software was pursued. When such solutions were not available or could not be used, customised solutions were developed.
\end{abstract}

\medskip


\section{Introduction}
\label{Introduction}

COMPASS is a fixed target experiment on the CERN SPS that uses tertiary high energy and high intensity muon or hadron beams with the aim of studying the nucleon spin structure and hadron spectroscopy. It is taking data since 2002, around seven months per year, and has shutdown periods in between, in which operations of maintenance and preparation of the following data taking period take place. The experimental setup is described in detail in \cite{COMPASS01}.

The detector devices and the experiment's environmental parameters are monitored and controlled using an experiment-wide DCS. This system must ensure a coherent, safe and efficient operation of the experiment, by providing clear and prompt information for the shift crew and detector experts in the COMPASS control room. Some complex subsystems of the experiment have dedicated stand-alone control systems. These systems communicate with the DCS, providing it the most relevant parameters. Since 2003, the COMPASS DCS has been an exclusive responsibility of the LIP-Lisbon group participating in the collaboration. This structure and organization is at contrast with the one of the big LHC experiments, which have a hierarchical structure, with distributed responsibilities~\cite{ATLAS,CMS,ALICE,LHCb}.

The DCS provides a graphical user interface for the shift crew in the COMPASS control room and detector experts to have access to all the relevant parameters monitored, their state (normal or in alert - indicated visually, by use of a color code, and acoustically) and their history, and a straightforward way to change their state, their settings, and the thresholds that define their state of alert.

\begin{figure*}
\centering
\includegraphics[width=0.8 \textwidth]{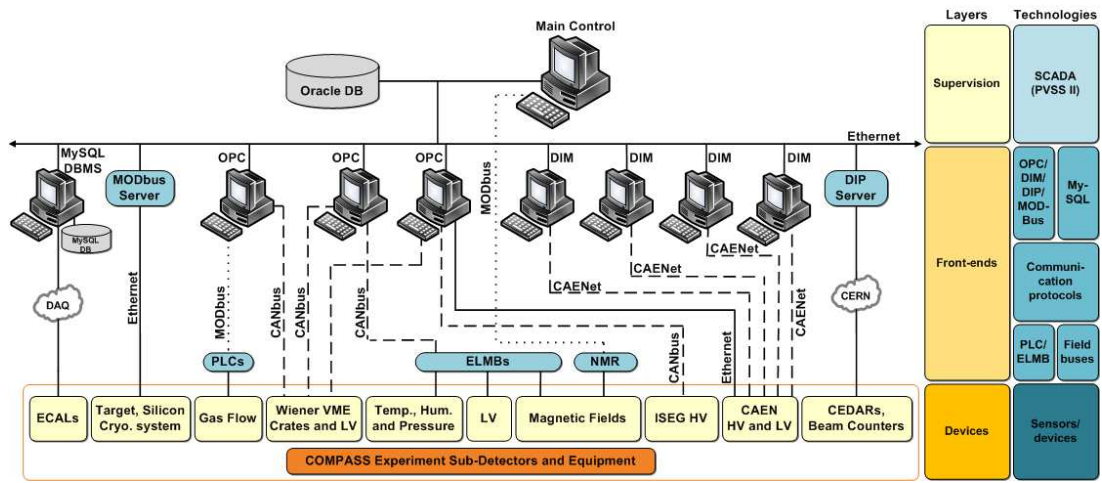}
\caption{The COMPASS Detector Control System architecture, comprising a devices layer, a front-ends layer and a supervision layer. The technologies used in each layer are indicated in the rightmost column.} 
\label{DCSscheme}
\end{figure*}

The architecture of the DCS is shown in Fig.~\ref{DCSscheme}. In the following sections, its different layers are described in detail: the supervisory layer, the front-ends layer and the devices layer.

\noindent\section{\uppercase{The Supervisory Layer}}
\label{Supervisory} 

On the supervisory layer, all the data collected, managed and published by the different kinds of servers or made available in databases is gathered, analysed and displayed to the end user. This includes visual and sound alarms in case of states of alarm. It also provides archiving of the data in an external database. Settings and alarm limits are also managed by the system.

PVSS-II\footnote{Pro\-zess\-teue\-rung und Pro\-zess\-vi\-sua\-li\-sie\-rung. See http://www.etm.com}, is the commercial SCADA system that was chosen by CERN to use in the LHC experiments, after a thorough evaluation process. Some of the aspects taken into account were: openess, scalability, cross-platform capability (i.e. to run in Windows and Linux) and long term support.

The COMPASS experiment has adopted PVSS early in its development phase and has been a benchmark for other experiments at CERN. In fact, before the LHC starting in 2010, COMPASS was the biggest experiment operating at CERN. Over the years, several versions of PVSS were used in COMPASS, that had been previously tested and validated both by CERN and by the COMPASS DCS group.
The installation of these patches during the data-taking periods requires a careful evaluation.

The JCOP Framework~\cite{JCOP} is a CERN project to develop common software tools for High Energy Physics related equipment and operations, to be used with PVSS. It provides templates of datapoint types, panels, functions and mass configuration tools for different classes of equipments or functionalities, providing, for instance, tools for the management of priviledges, or for trending plots.

The objects provided by PVSS and the JCOP Framework have sometimes to be adapted to meet COMPASS' needs. In addition, other solutions had to be developed independently for non-supported custom devices. This includes the control of custom devices,
accessed using their serial (RS232) interfaces or their web servers; 
or the monitoring of items from dedicated control systems, such as (EPICS, LabView, etc.\footnote{See http://www.aps.anl.gov/epics, http://www.ni.com/labview}), which are made available by various means, including mySQL and Oracle databases. 

The PVSS production system in use is both distributed and scattered. Historically, it started as a scattered project, meaning that it was constituted by a main PVSS project running in a Linux machine, and 3 associated PVSS projects running on Windows machines, that had PVSS processes running as OPC\footnote{OLE (Object Linking and Embedding) for Process Control. See http://www.opcfoundation.org} clients (7 clients in total). As the DCS developed, the main project was split into two distributed projects, for performance reasons. 

PVSS works with objects called datapoints, which are structures, {\it i.e.} they have a tree structure that can include branches and where the leaves are the the monitored and controlled parameters (and can have different types, such as floats, integers, booleans, strings, etc., or the corresponding array types). 

Presently, the project comprises over 20000 datapoints. Close to 17000 parameter values have alert handling, whereas almost 19000 parameters have their values archived. 

The polling rates are adapted to the rate of variation of the parameters, and range from one value per parameter read every 1.5 seconds (for fast varying parameters, sensitive to the beam, such as high voltage channels' actual values) to 2 minutes (for slowly varying parameters, such as high voltage channels' settings, or detector positions). For any given type of equipment, the items are grouped in PVSS subscription data sets according to these rates.

The access to the PVSS project is made available upon login. There is a general user name for the shift crew, user names for each of the detector experts, and a username for guests. For each login, there is an authorizations policy associated: certain operations are restricted (such as switching on or off the high voltage channels for guest users), or specifically allowed (such as saving recipes or reference files of high voltage settings; see 
details later in this text).

\begin{figure*}
\centering
\includegraphics[width=0.7 \textwidth]{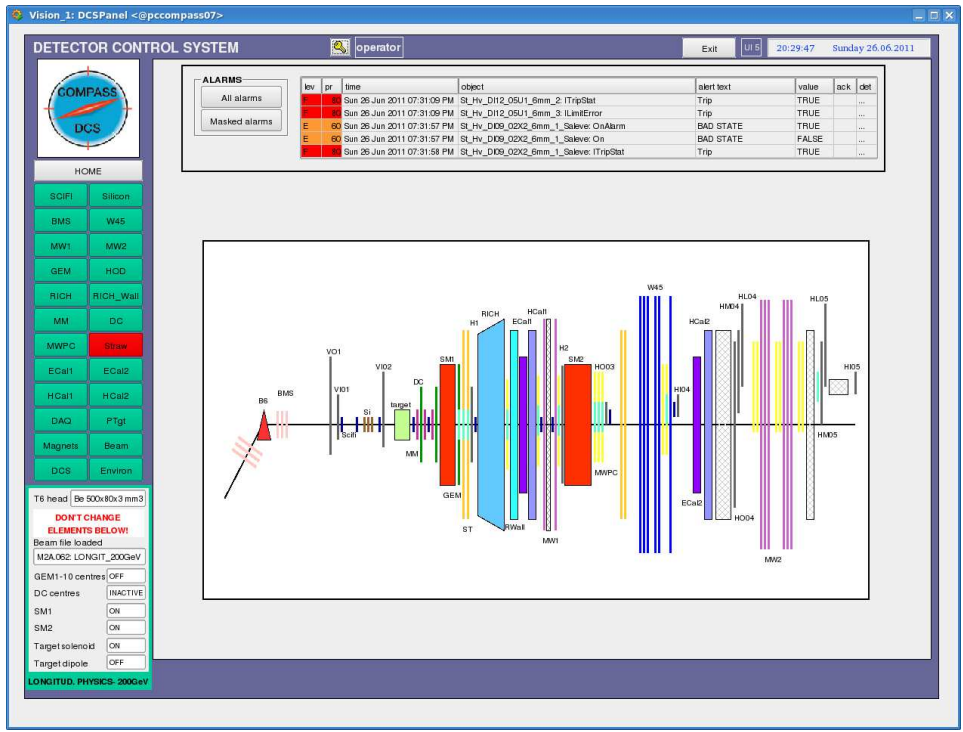}
\caption{The graphical user interface (UI) of the COMPASS DCS. One can see, on the top, the alert panel and, on the left, the buttons to access dedicated detector panels and, below them, a table with the summary status of the experiment.}
\label{ui}
\end{figure*}

The graphical user interface (UI) is the main mean for users to interact with the DCS. It is composed of multiple subpanels, organised in a hierarchical way, as can be seen in Fig.~\ref{ui}. One can see, on the top, the alert table and, on the left, the buttons to access dedicated detector panels and, below them, a table with the summary status of the experiment. In the larger area of the panel, a synoptic view of the spectrometer is displayed. This area is also used for navigation in the subsystems controlled and to display the actual data.

The datapoints history is made available online. In fact, PVSS trending plots (namely values over time) are one of the more useful and more used features of the DCS. In the user interface panels, customised buttons are created for the items that have numerical values (generally, floating point), so that their history can be easily accessed by users. The JCOP Framework provides its own trending plot widget, that was further customised for COMPASS, to make it simple to use. Users can choose the time range they want to visualise, change the scales or zoom in or out the abscissa or the ordinate by using the mouse scroll button, or choosing a rectangular region for zooming in by simply selecting two opposite vertices of the region. 
Templates can be created to allow, with one click, to see related parameters and adjust the ranges for each.
It is also possible for the users to include additional parameters to a predefined plot as well as to print the trends being displayed, or to save them to a file, not only as an image, but also in ascii format (comma separated values or CVS).

One of the most important functionalities of the DCS is the display of visual and audible alarms, when predefined conditions are met, namely, when parameter values beyond predefined thresholds are reached for datapoints with numerical values or when devices send alarm flags. The visual display of alarms follows a color code that indicates its severity. For the most relevant parameters, it is important to assure that the operator didn't fail to notice the alarm condition, even if it has disapeared in the meantime; in this case, it is requested that the alarms are acknowledged by the operator in the graphical user interface.
Upon activation of an alarm, detector experts are warned by email or SMS of states of alert in their detectors. 

Since the DCS has both a relatively fast knowledge of the state of the parameters it monitors and the ability to send commands to the devices, it is used to ensure software-wise protections to several equipments.
For instance, some detectors have components that are sensitive to magnetic field gradients; thus, when a trip of one of the spectrometer magnets (SM1 or SM2) occurs, or when these are switched off with the detectors high voltage channels still switched on, the DCS issues a switch-off command, so that the time interval during which the gradient is felt is minimised.
In addition, for some detectors, the high voltage channels should only be switched on or off in pairs; hence, if a trip is detected in only one member of the pair, the DCS sends a switch-off command to the other member of the pair.

Furthermore, some detectors have front-end cards that are refrigerated by a water circuit. When this circuit stops for some reason, the temperature of the cards increases and can reach a value above a predefined threshold. If this happens, the DCS issues a switch-off command to the low voltage power supplies that power them, thus preventing these sensitive and sophisticated cards to be damaged. A hardware interlock is activated at a higher temperature, but the recovery from the interlocked state implies access to the experimental area, and therefore represents greater beam time losses and is therefore to be avoided. 

A configurations database associated to the project was implemented. It is based in a JCOP Framework package, which was adapted for the COMPASS needs. It saves and retrieves data in an independent Oracle database. This database has two main purposes. On the one hand, it allows to save and to retrieve the so called ``recipes", {\it i.e.} sets of thresholds for alarms of groups of items. The recipes can be created, or its values committed in the PVSS project, using the DCS UI, provided that the user has the privileges to change the respective detector items.

The second important purpose of the configuration database is to store so cal\-led ``con\-fi\-gu\-ra\-tions", {\it i.e.} the mapping of hardware names vs.\ logical names ({\it i.e.} PVSS datapoint element names and respective aliases), for snapshots of stable states of the PVSS project. These are used to keep track of changes of the aforementioned mapping. These changes can happen either because, for instance, a high voltage channel gets broken and the same part of the detector ({\it i.e.} same alias) is then powered by a different channel ({\it i.e.} different datapoint), or because channels are reused when switching between the muon and the hadron Physics program data-taking.

For storage of settings of high voltage channels (set voltage, maximum current allowed, ramp up speed, ramp down speed and trip time), ascii files are used, for convenience. Experts can access the files, edit them, and send the values to be used by the equipment. The shift crew can use these reference file to recover the normal state of the equipment in case of problems.

Only a subset of all the data that PVSS receives and manages is actually saved. For this to happen, the PVSS datapoints need to have an archiving policy defined. This is chosen according to the known changes of each datapoint and the relevance of its history. For instance, it may be useful to store the readings of a temperature every ten minutes or if the change with respect to the previous reading exceeds one degree. This smoothing condition, called dead-band, is adjusted for each datapoint group or even per datapoint, if needed. The generic rates of archiving range from one value for every $\sim$40 seconds (corresponding to the beam supercycle time interval) for beam-related quantities, to one value for every half an hour (for the positions of detectors). 

Currently, there are around $2\times 10^9$ values stored ({\it i.e.} around 300 GB of data, including indexes), comprising the project history since 2006.
The DCS historical data that had been saved in PVSS internal format (during its first years of operation) was copied to a CERN central Oracle database, and the new data produced was all stored in this database. This way of storing data has all the advantages of Oracle and makes their access independent of PVSS. The data is continuously replicated to a second database, to ensure that the access to the data never compromises the performance of the production database. The data can also be provided in other formats, such as ascii or ROOT~\cite{ROOT} trees.

The DCS data is very important for studies of detector performances.
Some particularly relevant parameters for offline Physics data analysis are regularly copied to the experiment's mySQL conditions database, using a cron job.
The history of alerts of all the datapoint elements that have alert handling is also saved and made available by PVSS. This includes the timestamps of their arrival and departure, and of eventual acknowledgements done, among other information.

The knowledge of malfunctioning of parts of the experiment relies substantially on the DCS, namely on the display of alarms. Hence, it is important to assure its integrity and availability, ideally, at all times. Some of the mechanisms used are heart-beats, watch-dogs, back-ups, a security policy and the issuing of regular ping commands.

The managers of PVSS's main project, which are independent processes running in Linux, may, for a number of reasons, either block or stop running. On the other hand, the servers -- either OPC, \cite{DIM}, or other -- may stop delivering meaningful data. For this reason, for each manager in PVSS that acts as a client, a heart-beat item was created, that gives the timestamp of the last meaningful data it received.

Moreover, to verify that PLCs\footnote{Programmable logic controllers.} are sending meaningful data at all times, a mechanism of watchdogs is implemented. The OPC server marks as invalid values sent by the PLC in case the values of the items published for this purpose (which have, during normal operation, varying integer values), stop begin updated.

The communication with individual VME\footnote{VERSAmodule Eurocard bus.} crates or power supplies is also monitored, by continuously checking for selected equipment items that the timestamp of the latest value read is more recent than a predefined time interval.

To ensure the integrity of the project if a software corruption occurs in the PVSS project and associated software, the data is copied every twenty-four hours to a central repository. Furthermore, local copies are periodically made. 

A thorough security policy is implemented. All the computers that integrate the DCS belong to a dedicated experimental domain, that communicates with the CERN network using dedicated gateways. All the PCs in use have firewalls implemented.
In addition, all the user interfaces, with the exception of the one in the control room that should be permanently accessible, have an auto-logout after one hour.

The project should be available in the network at all times, for instance to diagnose of eventual DCS problems remotely. For this to happen, the gateways of the COMPASS domain have to be switched on and accessible via the CERN network. To check that this is the case at all times, regular ping commands are issued (every fifteen minutes) from an external server and the response is monitored; a notification is sent to the DCS experts in case those gateways are not reachable. 

\noindent\section{\uppercase{The Front-ends Layer}}
\label{FrontEnds}

The experiment devices that are monitored and controlled by the DCS are spread over nearly two hundred meters, including the spectrometer and the beam tunnel. To communicate with all the devices, different field buses and communication protocols are 
used, namely CAN\footnote{Controller area network. ISO standard 11898, see e.g.\ www.iso.org} bus (8 daisy-chains), CAENet (6 daisy-chains), ModBus\footnote{See http://www.modbus.org}, Profibus\footnote{See http://www.profibus.com} (4 daisy-chains) and Ethernet.The general baud rate used for monitoring in the COMPASS CAN buses is 125 kbaud ($\simeq 34$ kbits/s), which is the recommended baud rate for the length of the daisy-chains used. These field buses transmit the information about the measurements of sensors to the front-end PCs (and commands to actuators in the opposite direction). In the front-end PCs, standard PCI\footnote{Peripheral Component Interconnect.} cards are installed to collect the information carried by the field buses. The data is transmitted to the supervisory layer using a server-client model. An exception to this model is the three-layer model which is used when a database is included as an intermediate between the server and PVSS. This happens for the monitoring of the calorimeters, beam and trigger rates, and part of the polarised target system.

In ad\-di\-tion, spe\-cialised de\-vi\-ces are used as intermediates between the devices and some of the front-end PCs, namely ELMBs (Embedded Local Monitor Boards) and PLCs.

The ELMB, described in \cite{ELMB}, is a multi-purpose multiplexed ADC with 64-analog input channels with 16 bit-resolution which was developed by the ATLAS experiment. The communication of the ELMBs with the front-end PCs is done with the CAN field bus, using the CANopen protocol. The ELMB was designed and tested to be radiation- and magnetic field tolerant: its tolerance ranges up to about 5 Gy and $3\times 10^10$ neutrons/cm$^2$ for a period of 10 years and to a magnetic field up to 1.5 T.

The PLCs (Programmable Logic Controllers) are stand-alone, very robust, reliable and relatively fast devices that allow, among other operations, to regulate flows of gases and their percentage in mixtures, according to predefined settings and tolerance intervals, as well as to regulate cryogenic systems. The measurement of gas flows or gas percentages in mixtures is provided by the PLCs by ModBus to the DCS front-end PCs. 

Manufacturer's OPC servers are used when available and stable. This is the case for the modern CAEN equipment and for Iseg equipment. In order to communicate with PLCs, an OPC server from Schneider\footnote{See http://www.schneider-electric.com}, is used. Moreover, an OPC server was developed at CERN to control relatively old Wiener equipment, as the one used in COMPASS. To communicate with ELMBs, a CANopen OPC server, described in \cite{CANopen}, is used.

The Distributed Information Management system (DIM) was developed at CERN and allows the implementation of a server-client model of publishing of lists of items and their actual values. The SLiC\footnote{See http://j2eeps.cern.ch/wikis/display/EN/SLiC} DIM server developed at CERN, allows the control of the six CAENet lines used for the older type of CAEN crates. Each server has different groups of items with individually tunable speed reading cycles, thus permitting the separation of fast reading cycles (comprising voltages, currents and channel status) with reading frequencies as low as 1 Hz, thus allowing a fast detection of high voltage trips and failures; and of slow cycles, used for the read-back of settings.
				
The DIM protocol is also used to monitor other systems, namely temperatures and disk occupancy of servers, and processes of data transfer from the DAQ machines to CASTOR. 

DIP\footnote{See http://en-dep.web.cern.ch/en-dep/Groups/ICE/Services/DIP} is a protocol developed at CERN, based on DIM, but allowing exclusively read-only parameters, which are, in practical terms, those related to the CERN infrastructure (such as beam line magnet currents and the last beam file loaded, the primary target head inserted, the parameters to allow the monitoring of the CEDAR detectors, and data relative to the liquid nitrogen supply).

One PLC from the polarized target system is monitored using the S7 driver provided by PVSS, thus avoiding the use of an OPC server.

PVSS provides functions to access relational databases such as mySQL and Oracle. This allows the access of information from the experiment conditions database (a set of mySQL databases), such as the calorimeter calibration event amplitudes, the beam and trigger rates, and parameters related with the polarised target.

The high voltage system of some of the detectors (Micromegas and the so called Saclay Drift Chambers) have special requirements with regards to its monitoring, and thus have a dedicated control system based on EPICS. This system publishes the most relevant data, which is read by a specially developed PVSS API (Application Programming Interface).

The Profibus protocol is used to transmit the data coming from the PLCs that monitor the detector gas systems to the PC that runs the Schneidar OPC server.

Moreover, the magnetic field of the SM2 is measured with an NMRmeter that comes with a serial interface which, by use of the Profibus protocol, allows the communication with a standard PC, where a custom c program reads the information transmited, writes it in an ascii file and thereby makes it available for a PVSS API that collects the values and writes them into a datapoint.

\noindent\section{\uppercase{The Devices Layer}}
\label{Devices}

Many different types of devices need to be controlled or simply monitored by the DCS, from high and low voltage crates and VME crates, to gas systems, sensors of temperature, humidity, pressure and magnetic fields.

COMPASS uses CAEN crates of different models to power most of its high voltage channels and for part of its low voltage channels. About 20 CAEN crates of older models (with CAENet interface) are in use, and six crates of newer models (with Ethernet interface).
Seventeen Iseg high voltage modules are also in use and integrated in the DCS by use of their CAN interfaces. 
In addition, fourteen Wiener low voltage power supplies are controlled, of which four are of type UEP6000, eight of type PL6021 and two of type PL508L. Nineteen VME crates are integrated in the DCS, both of older models (power supplies of type UEP5021) and newer models (power supplies of type UEP6021), the former being the majority. Both the power supplies and the VMEs are controlled by use of their CAN interface.

In subsystems where PLCs are used, the DCS only monitors the values that are published by them. This happens for the detector gas systems, the CEDAR detectors, and for systems that have dedicated control systems, namely the cryogenic systems of the polarized target, liquid hydrogen target and cold silicon detectors, see \cite{Cesar, Anibus}.

A wide range of devices are monitored by use of the ELMBs.

Hundreds of sensors are installed to monitor specific components of detectors or the experimental hall environment. For temperature monitoring, PT100 sensors in a 4-wire configuration are extensively used, whose output is read using the ELMBs.

Some of the low voltage power supplies used in the experiment only have an interface for monitoring channel voltages or currents by means of voltage signals proportional to the values to be monitored, which is also read by ELMBs. In such cases, a calibration formula is used in the configuration of the CANopen OPC server, to provide the conversion to the real values delivered by the channels.

Moreover, two of the most important magnets of the experiment, SM1 and Bend6, have their magnetic field monitored by Hall probes, whose output signals are read by ELMBs.

In the case of the second dipole magnet of the experiment, an NMRmeter is used. The NMRmeter has a serial interface, which, by use of the Profibus protocol, allows the communication with a PC.

A custom power switch is also controlled, by use of the web server and driver provided with the equipment.

The DCS has an indirect monitoring of the powering system plus read-out chain of the four calorimeters of the experiment, based on the calibration signals of a laser system (for ECAL1) or a LED system (for the remaining three calorimeters). A component of the DAQ system calculates a spill-average amplitude of the signal read-out by each of the $\sim$4500 channels, see \cite{Konopka}, and saves this information in a mySQL conditions database, that is subsequently accessed by a PVSS script. In the DCS, a reference is chosen by the detector experts; afterwards, the DCS calculates, for each beam spill (using a synchronization scheme with the DAQ), the state of alert of each channel, based on the relative difference of the actual amplitude of the calibration amplitudes with respect to the reference values. The conditions to indicate alarms in the main panel are specified for the total of channels with a given alert state.

\begin{figure}[h]
\centering
\includegraphics[width=0.4 \textwidth]{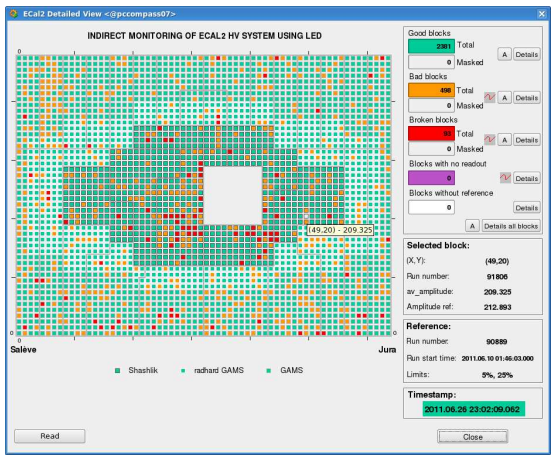}
\caption{Panel for monitoring of the high voltage system of ECAL2, comprising around 3000 channels.}
\label{ECAL2}
\end{figure}

The main power supplies powering the electromagnetic calorimeters, their monitoring systems, and the subgroups distributor voltages are monitored by use of ELMBs.

The positions and movement of the electromagnetic calorimeters are controlled by CAMAC\footnote{Computer Automated Measurement And Control.} modules, from whose readings the detector positions are calculated and read by the DCS.

A recent integration in the main PVSS project is the monitoring of the most relevant parameters of the complex Polarized Target system. The communication with the devices required the usage of different protocols and front-end solutions: PLC S7, ModBus, DIP and ODBC (for MySQL and Oracle database connection). 

The rates of the different triggers of the experiment, mo\-ni\-to\-red online by the shift crew, are stored in a mySQL conditions database read by the DCS, which calculates rates normalised to the beam flux, and triggers alarms when those normalised rates fall outside predefined ranges.

The servers used in the DAQ run DIM servers to publish data related to internal temperatures, occupancy of their disks and status of important processes.

The beam line M2 belongs to CERN's infrastructure and thus is monitored and controlled by dedicated programs. The most important parameters, such as magnet currets, collimator positions, the primary target head or the beam file loaded are made available via DIP, thus providing alarms and and historical values of these parameters.

Moreover, the CEDAR detectors (\v{C}Erenkov Differential counters with Achromatic Ring focus), used in the hadron program of the experiment, are a responsability of CERN, and its relevant parameters are published using a DIP server. For the operation of these detectors, the density of the gas used must be within a predefined range. When this doesn't happen, the DCS displays a state of alarm, allowing the shift crew to start a procedure to refill the detectors. The high voltage system and the motors are also monitored.

\noindent\section{\uppercase{Conclusions}}
\label{Conclusions}

The DCS of the COMPASS experiment at CERN was presented in detail. This is a centralized system that displays to the end user in a homogeneised graphical user interface many different subsystems that use very different devices and thus require the use of a wide range of front-end solutions.

\section*{Acknowledgements}

We gratefully acknowledge the Controls group of CERN (IT/CO and, later, EN/ICE) and CERN's PhyDB for their constant and efficient support. This work was supported by FCT.

\bibliographystyle{IEEEtran}
\bibliography{anunes_bib}
\end{document}